\documentclass[a4paper,12pt]{article}
\usepackage[cp1250]{inputenc}
\usepackage{czech}
\begin{document}
\begin{center}
  {\bf             Realistic theory of microscopic phenomena;   \\
                a new solution of hidden-variable problem.  } \\   [3mm]
                       Milo\v{s} V. Lokaj\'{\i}\v{c}ek          \\
{\footnotesize
   Institute of Physics, Academy of Sciences of the Czech Republic  \\
                         18221 Prague 8,   Czech Republic     }
\end{center}
\vspace{0.5cm}

{\footnotesize
 \bf Abstract}

{\footnotesize

    Einstein's critique of the Copenhagen interpretation of quantum mechanics
was based fully on logical and philosophical arguments,  while this
interpretation seemed to be strongly supported by von Neumann who argued that
any local (or hidden-variable) theory is not admissible by the
quantum-mechanical mathematical model. However, his statement was not true as 
shown
convincingly by J. Bell in 1964; an unphysical assumption was involved in von
Neumann's approach. After modifying the corresponding axiomatic basis Bell
derived some inequalitities, which was believed to hold for any kind of local
(hidden-variable) theories applied to EPR experiments in Bohm's
modification (coincidence measurement of spin orientations).  These 
inequalities
were then violated by experimental data, which was interpreted mostly as 
an important support of the Copenhagen interpretation. However, Einsteins
objections were not removed, and thus the question has remained: Do the
assumptions involved in Bell's approach cover really all possible
hidden-variable theories or not? After the experience with von Neumann's
proof such a question has been fully justified.  Doubts should concerned 
 the fact that  some non fully reasoned mathematical operations
(interchange of probability factors) was made used of in deriving 
Bell's inequalities.  These operations have been shown recently 
(http: //xxx.lanl.gov/quant-ph/9808005/) to be allowed only if an
important simplification in a general local theory has been
introduced. Similar a-priori assumptions have been involved in other
kinds of derivations of the given inequalities. 

   As to the quantum-mechanical mathematical model (representing the basis of 
all considerations)
the central problem is to be seen in the definition of the Hilbert space in
which all solutions of time-dependent solutions of Schroedinger equation have
been represented. The actual physical properties of such solutions have been
strongly modified if Hamiltonian eigenfunctions have been chosen as the
vector basis of the given Hilbert space and an (even basic) physical
meaning has been attributed to them. Having chosen another (suitably 
extended) vector basis it has been possible
to propose a realistic theory of microscopic phenomena  covering 
all experimental results that have been claimed to be in agreement 
with predictions of  
the standard quantum-mechanical model.  It concerns not only all
stationary characteristics of microscopic physical systems but also the EPR
experiments even if they must be interpreted rather differently now as the
internal space (geometrical) structure of measuring devices must be taken 
into account.
In addition to spin orientations they are also impact parameters in
collisions of photons with individual interaction centers that play the
role of hidden variables from the point of
view of the standard quantum mechanics. All previous logical problems have 
been removed and any paradoxical behavior need not be  required more in the 
proposed extended model.

---------------------------------------------------------------  } \\[5mm]

{\small    {\bf  1. Introduction   }

    The most physicists believe  in Copenhagen
interpretation of the quantum mechanics including paradoxical
behavior of the microscopic world, even if it has not been shown
until now how the laws of macroscopic  world might be derived from
such a basic microscopic concept. Consequently,  discussions concerning
the famous controversy between Einstein \cite{ein} and Bohr
\cite{bohr} have intensively continued in the whole century and corresponding
symposia have been held in regular intervals (see, e.g., \cite{phai}).

     In all these discussions  the standard quantum-mechanical mathematical
model has been practically always assumed to
represent the only possible description of microscopic
physical phenomena.  The discussions have  concentrated
practically  to the  question, which of the two different interpretations
of such a model should be preferred:

 (i) orthodox (Copenhagen) - all information about a physical system
at a given instant $t$ being contained in a  $\psi(x,t)$-function derived
by solving the corresponding time-dependent Schr\"{o}dinger equation
and represented by a vector in the Hilbert space spanned on Hamiltonian
eigenfunctions;

 (ii) ensemble (statistical) - only statistical characteristics being
described by such a $\psi(x,t)$-function, which corresponds
to Einstein's opinion that quantum mechanics cannot be considered
complete.

   The latter interpretation requires, of course, the existence of some
other parameters that would characterize a  physical system at
 a given time in addition to stationary quantities defining the usual
$\psi(x)$-function. That requires, however, to look for a
mathematical model in which the corresponding so called hidden
variables will be included. However, practically until now any actual serious
attempt of proposing  a corresponding extended model has not been done (or at 
least has not been successful).

    Such an extended model and its physical characteristics and consequences
will be described in the following. However, to make the problem more
understandable it is necessary  to  start
with the analysis  of the contemporary  quantum-mechanical model.  In
individual sections the following  items  will be treated:

 (i) the source of paradoxical properties in the standard
quantum-mechanical model will be discussed;

 (ii) an extended model will be proposed
and its properties will be shown;

 (iii) the actual meaning of EPR experiments (and corresponding
experimental results) will be discussed in the light of the new extended
model;

 (iv) some predictions of the extended model (differing
from those of the standard quantum-mechanical model) will be
mentioned;

  (v) the way how  the deterministic (semiclassical) behavior and the
probabilistic one  combine in the extended model will be discussed;

  (vi)  consequences concerning some physical concepts as well as
general thinking of human society will be mentioned.     \\

 {\bf  2. Quantum mechanics and the origin of paradoxical properties  }

   It is possible to state that the  following assumptions represent
basic ingredients of the  quantum-mechanical model:

  (i)  the behavior of any physical system is assumed to be described
by a  $\psi(x,t)$-function obtained as a solution of
time-dependent Schr\"{o}dinger equation
\begin{equation}
     i\frac{\partial}{\partial t}\psi(x_j;t) = H \psi(x_j;t)   \label{schr}
\end{equation}
where $H$ is the corresponding Hamiltonian and $x_j$ represent
space coordinates of all involved mass objects;

  (ii)  individual functions $\phi_\tau(x_j) \equiv \psi(x_j; \tau)$  are
 represented by vectors in an Hilbert space; all physical characteristics
being derived with the help of rules holding for operators and vectors in the
given Hilbert space, if the vectors represent individual physical
states and the
operators physical quantities  (see, e.g., Ref. \cite{gal}).

     However, these two basic assumptions would not have to lead to
any paradoxes (as will be shown later)
if  the structure of corresponding Hilbert space were not
defined with the help of the  additional assumption:

 (iii) the vector basis of the given Hilbert space is formed by
Hamiltonian eigenfunctions  $\psi_E(x_j)$:
\begin{equation}
     H\psi_E(x_j)  \; = \; E \psi_E(x_j)  \; ,
\end{equation}
and a basic physical meaning is attributed to these
eigenfunctions. In addition to, the superposition principle has been
introduced and each unit vector of such a Hilbert space has been assumed to
correspond to a possible  state of a given physical system.

   We will show now that a kind of discrepancy exists between
assumption (iii) and  assumption (i). Schr\"{o}dinger equation
(describing a physical system defined by corresponding
Hamiltonian $H$) provides different solutions characterized by a set
$\kappa$ of physical quantities being conserved during whole evolution
and by initial conditions of changing quantities (i.e., before all by
positions and momenta of individual matter objects). The values of all
these quantities are defined as expectation values
of corresponding operators.  It is possible to derive time-dependent 
positions and momenta of all objects from individual solutions $\psi(x_j;t)$, 
 i.e., as the expectation values of corresponding operators,  which
may be brought to the correspondence with solutions of Hamilton equations
(describing the  system consisting of $N$ matter objects with the help of 
thw same Hamiltonian). Consequently, they might be
represented in principle by  corresponding trajectories in the standard
($6N$-dimensional) phase space; individual
trajectories being mutually fully separated (i.e., without any common points).
 And it might be expected that evolution of a physical system (consisting
of $N$ stable matter objects) will be represented by trajectories with
similar properties also in  Hilbert space chosen in a suitable way in
harmony with assumption (ii).

   However, the situation have changed drastically when assumption (iii) has
been added to the first two assumptions,
and especially, when Hamilton eigenfunctions have corresponded to
basic physical states.
 Individual solutions of Schr\"{o}dinger equation  (\ref{schr})
may be then represented by  sequences of unit vectors in such an Hilbert
space or by trajectories crossing mutually; i.e., states belonging
to differently chosen
initial function $\psi(x_j;0)$ or to different values of $\kappa$ are often
represented by the same vectors. In such a case
a special measurement postulate had to be introduced
\cite{neu} enabling to derive predictions concerning experimental data.
 And one can say  that practically all quantum-mechanical paradoxes follow
from  assumption (iii).

   Representing  the solutions of Schr\"{o}dinger equation
in an Hilbert space has provided, however, a  suitable basis for
analyzing microscopic phenomena when some processes must be described as
probabilistic. And one must ask if it is possible to modify assumption (iii)
to be in better agreement with the characteristics of Schr\"{o}dinger
equation.  The possibility of using a more suitably defined (extended)
Hilbert space was discussed, e.g., by
 Rosenbaum \cite{ros}  in 1969 (and subsequently by some other
authors). However, the extension considered in \cite{ros} is to be denoted as
 too formal and too general; it did not
reflect actual physical conditions. In the next section a kind of minimum
extension reflecting the physical situation will be described
and some physical consequences derived.   \\

 {\bf  3. Extended Hilbert space  }

    A suitable extension of the Hilbert space was proposed in principle
independently  by three groups of  authors many years ago
\cite{lax,alda,newt}.  They were Lax and Phillips \cite{lax} in 1967
who defined such a mathematical model for the first time; they used it for
the description of some acoustic and optical phenomena, which invoked
impression that the model was suitable for being applied to some
semiclassical problems only. Later the same Hilbert structure
was derived by Alda et al. \cite{alda} in solving the problem of a
purely exponential decay law of unstable particles. And finally,
 Newton \cite{newt}  showed that  it
was possible to define regularly the time operator in the case of
harmonic oscillator when a similar extended Hilbert structure was
made use of, while it was not possible to do it in the standard
Hilbert space as pointed to by Pauli in 1938.

    We shall illustrate basic characteristics of the mentioned 
approach with the help of the
physical system consisting of two zero-spin particles, which represents the
smallest system exhibiting time evolution.   Its behavior in the
center-of-mass system may be described by  Schr\"{o}dinger  equation
(\ref{schr}) with  Hamiltonian
\begin{equation}
       H  =   \frac{p_j^2}{2m} +  V(q_j)   ;
\end{equation}
where $m$ is the reduced mass of the  particle pair, and the
operators of relative coordinates $q_j$ and of momentum components $p_j$
of
one particle  (in the center-of-mass system) are assumed to fulfill the
following  commutation relations
\begin{equation}
      [q_j,p_k] = i\delta_{jk}, \;\;\;   [p_j,p_k] = 0 ,
                                              \;\;\; [q_j,q_k] = 0 .
\end{equation}

    Introducing two other operators
 \begin{equation}
        Q \; =\; q_j^2 ,  \;\;\;\;\;  R \; = \; \frac{1}{2}\{p_j,q_j\}
\end{equation}
and assuming   $V(q_j) = V(Q)$  (i.e., the mutual potential depends on the
distance between particles) one can write further
\begin{equation}
 i[H,q_k]\; = \; \frac{p_k}{m} ,   \;\;\;\;
              \;\;\; i[H,p_k] \;=\; -2q_k\frac{dV(Q)}{dQ} \; , \;\;
\end{equation}
\begin{equation}
   i [H,  Q] \; = \;\frac{2}{m} R ,  \;\;\;
         \;\;\;\;\;  i [H, R] \;= \;2\{H-V-Q\frac{dV(Q)}{dQ} \} \; .
    \label{rh}
\end{equation}
 It is also possible to introduce  angular-momentum operator
fulfilling  relations
\begin{equation}
      M_{jk} \; = \; [p_j,q_k] ,   \;\;\;\;\;  [M_{jk}, H ]\; = \; 0 ;
\end{equation}
   and further  operator
\begin{equation}
     M  \; = \; M_{jk}M_{jk}
\end{equation}
fulfilling relations
\begin{equation}
            [M,M_{jk}] \; = \;     [M, H]  \; = \; 0 .
\end{equation}

       Individual trajectories corresponding to different solutions of
Schr\"{o}dinger equation may be then  characterized by expectation values of
mutually commuting operators; i.e., by $<H>$, $<M>$,
and,  e.g., by $<M_{12}>$,  which may represent the mentioned $\kappa$ set for
a given particle pair.  Different points on individual trajectories may be
then distinguished with the help of expectation values of operator $R$. 

    Assuming the particle pair to be in an unbound state (i.e., belonging
to continuous energy spectrum of the Hamiltonian) it follows
from Eq. (\ref{rh}) that the  expectation value of $R$
always increases for smooth repulsive potentials; i.e., when the function
$dV(Q)/dQ < 0$. It may rise, of course, for attractive potentials,
too, as far as the kinetic energy is sufficiently large;
the expectation value of it  increases
in principle from $-\infty$ to $+\infty$. Negative values of $<R>$  characterize
incoming states of the particle pair and positive values  outgoing states; minimum 
distance between both the particles corresponding to $<R> = 0$.

    For the states belonging to the discrete part of Hamiltonian spectrum
(particle pair being in a bound state) the  operator $R$ ceases to exhibit a
monotone behavior; it changes periodically during the evolution. In
such a case its expectation values are not sufficient to distinguish between
all different physical states and some other operators must be introduced, as
will be shown in the other part of this section (the case of  harmonic oscillator).

    In any case the standard Hilbert space (defined according to
assumption (iii))  is not sufficient to
characterize  behavior of a particle pair in agreement with
reality as all states belonging to different values of $R$ are
represented practically by one common vector.   To distinguish
all different states (e.g., pair particles at different distances)
the mentioned extended Hilbert space must be
made use of. As to the continuous Hamiltonian spectrum it is the
Hilbert space introduced in Refs. \cite{lax,alda}. For bound states
(corresponding to discrete spectrum) the structure of Hilbert space
is a little different; see the example of
harmonic oscillator (comp. also Ref. \cite{newt}).      \\

 {\it Continuous  Hamiltonian spectrum }

   We will discuss  the special case of conformal potential
\begin{equation}
                             V(Q) \; = \;  \eta Q^{-1}\; .    \label{pot}
\end{equation}
 The corresponding Hamiltonian has  continuous spectrum for
all real values of $\eta$, and only collision states exist.
 It holds then in such a case
\begin{equation}
                            i[H, R] \; = \; 2H   \;
\end{equation}
and the operator $R$ exhibits evidently a constant increase for any (positive)
energy, independently of $\eta$ value.   It is possible to define  operator
\begin{equation}
                   T \; = \; \frac{1}{4} \{H^{-1}, R\}  \label{tc}
\end{equation}
 fulfilling the  commutation relation:
\begin{equation}
        i[H,T] \; =\; 1 .        \label{tim}
\end{equation}
It means that it is possible to define the time operator as a function of
 operators $q_j$ and $p_j$.  One-to-one correspondence exists between
expectation values of $R$ and $T$; zero values for both these
operators corresponding to the minimum distance between the two particles.

    It holds  for expectation values of $T$ corresponding
to instantaneous states $\psi_\kappa(x_j;\tau)$ of a particle-pair system:
\begin{equation}
   <\psi_\kappa(x_j;\tau)| T |\psi_\kappa(x_j;\tau)> \; = \; \tau
\end{equation}
for any $\kappa$ and $\tau$. Introducing the evolution operator
\begin{equation}
      U(t) \; = \;  e^{-iHt}   \hspace*{1cm} (t > 0)   \label{ut}
\end{equation}
it holds also
\begin{equation}
      U(t)\; |\psi_\kappa(x_j;\tau)> \;=\; |\psi_\kappa(x_j;\tau+t)> \; ,
\end{equation}
which indicates that expectation values of the operator T defined by Eq.
(\ref{tc}) may be hardly identified with the parameter of flowing time.
They characterize instantaneous states in a special scale, i.e., with the
help of time expressing the distance from the state $\psi_\kappa(x_j;0)$. The
evolution operator  moves the states always to higher values of $\tau$,  or
from in-states to  out-states (belonging to negative, resp. positive,
expectation values of $T$).

   As to the structure of the extended Hilbert space  we
have already mentioned that it was introduced by Lax and
Phillips \cite{lax} and derived in Ref. \cite{alda} by requiring an exact
exponential decay law to hold for unstable objects.  A detailed mathematical
description  of such an Hilbert space may be found also in Ref.
\cite{lax2}.
 Taking a  $\psi(x_j)$ function (i.e., for a special value of $t$) from
possible solutions of  time-dependent Schr\"{o}dinger equation
one can easily see that the same function may belong to an
incoming state as well as an outgoing one. To distinguish such two
states it is necessary for  the corresponding Hilbert space to consist of two
mutually orthogonal subspaces (corresponding to incoming and outgoing
states); the bases of each of them being formed by all
functions $\psi(x_j)$ derived by solving the
time-dependent Schr\"{o}dinger equation; i.e. by all  functions
$\phi_{\tau,\kappa}(x_j) \equiv \psi_\kappa(x_j;\tau)$
 belonging  to all possible values of $\kappa$ and $\tau$.
 The basis of the whole Hilbert space must be then
defined with the help of function pairs; i.e. by
\begin{equation}
 \Phi_{\kappa,\tau}(x_j) \; \equiv \; {\large\{}\frac{1-
\varepsilon(\tau)}{2}  .\phi_{\tau,\kappa}(x_j), \;
\frac{1+\varepsilon(\tau)}{2}
                   .\phi _{\tau,\kappa}(x_j) {\large \}}       \label{doub}
\end{equation}
where
\begin{equation}
         \varepsilon(\tau) \;  =  \; {\mathrm sign}\; \tau
\end{equation}
and $\kappa$ represents the set of expectation values  $<H>, \;<M>$  and
$<M_{12}>$.

       The solutions of the time-dependent Schr\"{o}dinger equation for
the conformal potential may be found e.g. in Ref. \cite{jack}. We will limit
here  to the special case of  $\eta = 0$;   the general structure of the Hilbert
space being  fully conserved.  One can write then in Eq. (\ref{doub})
\begin{equation}
  \phi _{\tau,\kappa}({\bar x})  \; = \;
   \int d{\bar k}\; g_\kappa({\bar k})\; e^{i{\bar k}({\bar x}-{\bar x}_i
                 -{\bar k}\frac{\tau-\tau_i}{2m}) }     \label{ini}
\end{equation}
where ${\bar x}_i$ and $\tau_i$ represent  values corresponding to
an initial state and
 function $g_\kappa({\bar k})$ fulfils the condition
\begin{equation}
        (2\pi)^{-3}\int d{\bar k}\mid g_\kappa({\bar k})\mid^2 \;=\; 1
\end{equation}
is an arbitrary function of vector ${\bar k}$, at least in principle; it must be
chosen so as to correspond to  values $\kappa$  in an initial three-dimensional 
state.

    The states in the extended model are characterized also by expectation
values of operator  $M_{jk}$, which enables to establish immediately
impact parameter for two-particle collision states. It may be defined as a 
position vector corresponding to the  minimum distance
between the particle pair during the evolution (i.e., in the  state 
characterized by  $<R>\; =\; <T>\; = 0$).  The impact
parameter value   may be
derived from $\kappa$ set and represents an indivisible part of
characteristics belonging to the extended model.  If two extended objects
collide important changes of the physical system may occur in transitions
from in- to out-states in dependence on impact parameter value, which is,
however, the problem lying outside the scope of this paper.
In the standard quantum-mechanics the impact parameter has 
been eliminated from the description by the $\psi(x)$-function and its 
value (or its probability distribution) has had to be added. In the extended 
model the statistical distribution of  initial value  $\bar x_i$ is responsible 
for probabilistic characteristics of the quantum-mechanical measurements.    \\

 {\it Discrete Hamiltonian spectrum}

   It is then possible to represent internal evolution of a bound
particle pair in a correspondingly extended Hilbert space in a similar 
way. There
are, of course, some differences against the preceding case as the
expectation value of $R$ changes periodically and it is not
possible to make use of it in defining individual states of a
bound system. It is necessary to introduce more suitable
operators.

   We will demonstrate the corresponding approach on the example of
harmonic oscillator when the Hamiltonian possesses a mere discrete
spectrum; i.e., only bound states may exist.  The potential between two
particles may be written now as
\begin{equation}
                       V(Q) \; = \; \frac{k}{2}\; Q  \; .
\end{equation}
The attractive force aims always to the common center of mass
and the behavior of a
three-dimensional harmonic oscillator may be described as the  product of
three (or at least two) linear harmonic oscillators.  In the following we
will limit ourselves to the simple one-dimensional case.

    Instantaneous states of  linear oscillator may be  described
with the help of operators
\begin{eqnarray}
     C &=& \sqrt{\frac{k}{2}} \left\{ H^{-1/2},q \right\},\\
     S &=& -\sqrt{\frac{1}{2m}} \left\{ H^{-1/2},p \right\},
\end{eqnarray}
which fulfill relations
\begin{eqnarray}
          i \left[ H,S \right] &=& \omega C , \\
          i \left[ H,C \right] &=& -\omega S ,  \\
            < C^2+S^2 > &=& 1  ,  \\
            <  CS-SC > &=&   0
\end{eqnarray}
where
\begin{equation}
          \omega = \sqrt{\frac{k}{m}} \; .
\end{equation}
   It is then possible to introduce the phase operator
\begin{equation}
  \Phi \;   =\;  arccos\; C       \;        = arcsin\; S
\end{equation}
and also the time operator
\begin{equation}
   T  =  \frac{1}{\omega}\Phi  \;    \label{th}
\end{equation}
fulfilling Eq. (\ref{tim}).

     The representation Hilbert space is now more complicated than in the
preceding continuous case. While  physically different states
are distinguished with the help of expectation values of $C$ and
$S$  the expectation values of $\Phi$ and $T$ may be equal to any
real value as evolution operator (\ref{ut}) evokes their steady
increase. Two mutually orthogonal subspaces should then correspond
to any interval $(2n\pi, 2(n+1)\pi)$ of expectation values of phase
operator $\Phi$.  The evolution operator  moves the states  from one
subspace to another;  physically important values $<C>$ and $<S>$
changing periodically.  Details concerning this Hilbert space
structure for harmonic oscillator will be given elsewhere.

    As to the given structure it is possible to say that it may be denoted as  a
solution of the  phase-operator  problem  opened  already by Dirac in 1927
\cite{dir} and not yet satisfactorily solved (see, e.g., Lynch
\cite{lyn}). Very recently Ozawa \cite{oza} showed that the problem
might be solved in the framework of an extended Hilbert space
constructed by him to such a purpose.
It might be interesting to compare the structures of the Hilbert space of
Ozawa and that of ours.   \\

{\it  Semiclassical properties of the extended model}

   The proposed extended model enables to describe the behavior of
microscopic physical systems consisting of a fixed number of stable
particles practically in a semiclassical and fully realistic way.
The evolution is  represented by a trajectory in a Hilbert space
characterized by  corresponding values  of  $\kappa$.
Consequently, any additional measurement postulate is not more
needed. There is not any difference in the predictions of
the stationary characteristics (being conserved during whole evolution)
by the standard quantum mechanics and by the
extended model.

   Important difference concerns the description of dynamic
processes as there is not more possible for states belonging to different
stationary characteristics to combine (and to form new pure states by
superpositions); any transitions between different $\kappa$ values are
not possible, either. They are also Hamiltonian eigenfunctions that do not
belong to extended Hilbert space.
The proposed extended model enables then to describe newly
transition phenomena, i.e., inelastic collisions and decay
processes (see Sec. 6).  Also the EPR problem may be now discussed from a
quite new point of view.   \\

{\bf   4.  EPR experiments and Bell's inequalities  }

      The original goal of the  EPR Gedankenexperiment proposed by
Einstein et al. \cite{ein} was to argue on logical grounds that
quantum mechanics could not be considered complete, and consequently, that
it was necessary to add some other characteristics (i.e., the so called
hidden variables) if any microscopic system is  to be described fully and in
a realistic way. However, the logical
arguments were not sufficient for the then physical community. The
most physicists believed in the standard mathematical model more
than in a realistic interpretation.  And Bohr's arguments  \cite{bohr}
were almost generally accepted at that time.

    Nevertheless, the discussion concerning the controversy between
Einstein and Bohr has continued during this century. It is not
possible to repeat the whole story of this problem here.  We shall
start with the impact that came  when J. Bell \cite{bell} derived his
inequalities. It was hoped  that it would be possible to solve  the old
controversy with the help of experimental results on their
basis.  A new search for feasible experiments of EPR type was
initiated.

      Experiments based on the coincidence measurements
of two equally polarized
photons passing through polarizers in opposite directions were
proposed and also performed. If  $a_\alpha$ and $b_\beta$ are probabilities
of single photons passing through individual (opposite) polarizers at given
settings $\alpha$ and $\beta$ then it should hold according to Bell in the 
realistic interpretation for
the combination of any four coincidence probabilities:
\begin{equation}
      a_{\alpha}b_{\beta}\; + \; a_{\alpha'}b_{\beta}\; +
         \; a_{\alpha}b_{\beta'}\; - \; a_{\alpha'}b_{\beta'} \;\le \; 2  \; .
        \label{bel}
\end{equation}

   The series of corresponding experiments started approximately in 1971 and
were finished practically in 1982 (see Ref. \cite{asp}) with the following
results:

 (i)  Bell's inequalities (\ref{bel}) have been surely violated for specially
chosen  orientations of polarizer axes, for which according to
quantum-mechanical predictions a value greater than two should be obtained;

(ii) the experimental results may be regarded as being in agreement with
quantum-mechanical predictions.

     These results seemed to prove definitely the quantum-mechanical concept
of particle non-locality. And a series  of physicists started to develop world
picture based on such a concept. Discussions were initiated how to make use
of the concept of non-locality in practical applications;  see a series of articles
published  in Physics World \cite{world}  about the so called
teleportation, cryptography and so on.

       However, all theoretical considerations concerning these problems have
been based on some additional assumptions going often far beyond the
standard quantum mechanics; sometimes even  contradicting them; see Ref.
\cite{lok3}. And all experimental results being claimed as a support of these
new ideas have been interpreted incorrectly as argued recently also by
Klyshko \cite{kly}.  Sometimes the given interpretation has been based on the
reversal of logical implication. There is not any doubt, either, that the so called 
interference phenomena (i.e., 
periodically changing results in all
these experiments) are based on changing  time of flight of individual
photons between two macroscopic objects and should be interpreted on the
same basis as Newton fringes  as mentioned
also  in  Ref. \cite{lok3}.

    There are, however, also papers looking for a more realistic concept of
the physical world. E.g., S. Goldstein \cite{gold}  returned recently to
Bell's ideas starting from Bohm's results,  which made it possible to
attribute  a definite track  and to go back to an ontological
interpretation of the
microscopic world. However, such a goal can be hardly reached if
one tries to solve the problems in the framework of the standard
quantum-mechanical mathematical model.
 The ontological tendencies are, of course, in a full harmony with the
just discussed extended model
the  principles of which were mentioned for the first time in Ref.
\cite{kazi}  and more systematically explained in Ref. \cite{lok1}.

      It is the violation of Bell's inequalities by experimental data which
still seems to represent an important argument in favor of  the standard
quantum-mechanical model and, consequently, against the extended model.
However, it has been shown recently that these inequalities have been
applied to experimental data in an inadequate way.

     It has been always believed that Bell inequalities have been fully
based on locality condition only. However, already in the first derivation
Bell had to use an additional mathematical operation (interchange of some
factors belonging to different coincidence combinations),  which was
regarded as fully acceptable. In fact this operation has represented a step
by which the locality concept has been significantly limited. And
similar limiting assumption has been used in all other kinds of deriving these
inequalities, the problem having been analyzed in Ref.  \cite{lok2}.

  While the space orientation of photon spin has been taken always into
account the additional assumptions (corresponding to used approaches)
has allowed to respect neither an actual space orientation of the photon
pair nor any internal microscopic structure of polarizers (or measuring
devices).  Then it is not possible to respect
 exact impact parameters of individual photons into the atom grids of
 polarizers at  given settings (and their corresponding weight
distributions); any averaging in coincidence experiments must be 
performed over given pairs
(not over events in single polarizers).  Bell inequalities cannot be derived
when a full realistic (locality) concept is taken into account.
 And it is not more possible to argue that the  results of EPR 
experiments contradict the locality of microparticles \cite{lok2}.

   Then of course, one point more remains to be explained: the fact that
EPR results may be taken  as being in agreement
with quantum-mechanical predictions. In this case it is
necessary to start  with
explaining the actual essence of EPR experiments. Their result does not
concern any direct predictions of quantum-mechanical model. The
experiments consist in demonstrating that the  results are the same in
one-side arrangement as well as in the coincidence  one:
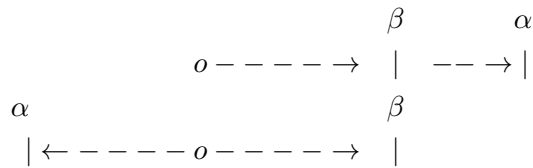
\begin{eqnarray}
         &\beta& \hspace{1.1cm} \alpha      \nonumber \\
   o----\rightarrow & |& --\rightarrow  |   \nonumber       \\
     \alpha \hspace{4.4cm} &\beta&                    \nonumber  \\
 | \leftarrow ----o----\rightarrow & | &      \nonumber
\end{eqnarray} \\
    I.e., in showing that the  transfer of non-polarized light through
two polarizers may be described in both the cases
with the help of the generalized Malus law ($\beta$ being put zero)
\begin{equation}
    M(\alpha) \; =\; (1-\varepsilon)cos^2\alpha \; +\; \varepsilon
     \label{gmal}
\end{equation}
where $\varepsilon$ is always a non-zero quantity. To obtain identical
results in both the different arrangements follows immediately from the
quantum mechanics for ideal polarizers (when $\varepsilon = 0$). And it is
assumed (without any actual proof)
that the same  may be expected for real polarizers, too.

    One can, however, show that the same predictions  may be derived for both
the arrangements in  a very simple local (hidden-variable)
theory. Assuming for simplicity (in the first approximation)
that the change of photon polarization during
its passage through a polarizer may be neglected, one can write in both
the arrangements
\begin{equation}
    M(\alpha) \;=\; \int d\lambda \; p_1(\lambda)\; p_2(\lambda-\alpha)
    \label{mal}
\end{equation}
 where $p_j(\lambda)$ are transfer probabilities of the light (photons)
through individual polarizers; $\lambda$ representing the deviation of the
polarization direction from the axis of  the first polarizer (putting again
$\beta = 0$).

    Consequently, in contradistinction to common opinion the results of EPR
experiments with polarized photons
cannot bring any decision  concerning the preference between the
standard quantum mechanics and the new extended  realistic
model.  However, the decision between these two  models may
be given with the help of other experiments, e.g., of those concerning the
light transfer through three polarizers.  \\


{\bf  5. Experiments with three polarizers  }

   As shown in the preceding the EPR experiments can hardly bring any decision
between the predictions of  quantum-mechanics and those of a hidden
variable theory. And we may ask, whether
 it is possible to find an experiment which could contribute to the
solution of this question. Combining Eqs. (\ref{gmal}) and (\ref{mal})
and putting $p_j(\lambda) = p(\lambda)$ it has been possible to determine
the shape of the function  $p(\lambda)$ when the generalized Malus law is to
hold for a  pair of polarizers.
And having used this  function in deriving
 angle dependence of light transfer through three polarizers
the predictions differing significantly from quantum-mechanical ones
have been obtained. The function  $p(\lambda)$ differs rather significantly 
from $cos^2(\lambda)$-function used by Belifante \cite{belf} in arguing 
that the predictions of a local theory are significantly different from the 
quantum-mechanical predictions.

    It means that a detailed analysis based on experiments with three polarizers 
could bring the whole problem by an important step further; especially, 
where preliminary measurements   (see Refs.\cite{kra1,kra2}) 
indicate that the results are rather far from any quantum-mechanical
characteristics. Predictions  corresponding  approximately to experimental
results  may be obtained with the
help of M\"{u}ller matrices starting from the description of the polarized
light proposed by Stokes. However, neither M\"{u}ller matrices seem to
be able to reproduce fully actual experimental results. Thus, the given
experiments  open new deeper questions,  whether commonly used mathematical
approaches are sufficient to characterize
 different degrees of polarization; the problem being discussed recently
also by Movilla et al. \cite{pol}.       \\

{\bf  6.  Deterministic and probabilistic behaviors of physical
systems and realistic model }

    Let us go now back to a physical system consisting of two
non-bound particles (states corresponding to continuous Hamiltonian
spectrum). As mentioned above the individual subspaces in the
extended representation Hilbert space may be denoted as subspaces of
incoming and outgoing states. And one can aver that any evolution
of a physical system should be regarded as  irreversible since
the evolution operator (\ref{ut}) transforms always the states from
negative $\tau$ to positive $\tau$   and never in the opposite
direction (i.e., from "in" to "out"). Even inside the individual
subspaces the evolution goes always from lower values to greater
values of $\;\tau$ (see \cite{alda,lax2}).

    Some more complicated behavior may occur around the
value of $\tau = 0$. In the case of an elastic collision the
evolution goes continuously from "in"- to "out"-states, being represented
by Eq. (\ref{doub}). However, if the mutual impact parameter is
sufficiently small then an inelastic collision of non-point objects 
may occur and the
characteristics of the system may change substantially. In such a
case  the system may pass to an out-state belonging to a quite
different type of particles. However, the following evolution goes
 again (in the corresponding out-subspace) along the trajectory
characterized by expectation values being conserved during the whole
evolution (i.e., by the set of given $\kappa$ values). Some
additional hidden parameters describing internal structures of
colliding particles and their instantaneous values  may play, of course,
an important role in such transitions.

      At the present we are forced to describe the transitions from
a given in-space to a different out-space with the help
of phenomenological probability functions  derived from
experimental data. In some cases  probability values may be
predicted on the basis of  stripping theory (e.g., in the case of
nuclear collisions).  In the case of hadrons the extended
model represents a new challenge to look for a corresponding more
realistic model of internal
characteristics of these matter objects and of their interactions;
the most hitherto hadron
collision theories being very far from a considered realistic concept.

   It is possible to conclude that the deterministic as well as probabilistic
behaviors are described  in the framework of one mathematical
model. The extended model enables to describe the deterministic
evolution of  semiclassical systems (i.e.,
when the numbers and kinds of objects do not change) in the framework
of individual subspaces of the total Hilbert space.  Probabilistic
processes (e.g., inelastic collisions) may be then characterized with the
help of transition probabilities from one subspace to  another one.

   The extended mathematical model opens the possibility to describe
in a realistic way not only
inelastic collisions  but also  spontaneous decay of
unstable particles (see Refs. \cite{alda,lax2}).  However, an
unstable particle may be  hardly represented by one vector in the
Hilbert space but by a subspace being orthogonal to all other parts
of the total Hilbert space. In the first approximation this
subspace might be taken as $(n+1)$-dimensional when the given
particle decays into $n$ different channels.  Such a structure of
an unstable particle  was considered already in Ref.  \cite{lok8}
(see also \cite{lok9}), in solving a kind of deviations from the
simple Breit-Wigner formula in resonance scattering. The unstable
(resonance) particle has been assumed to exhibit some random
transitions between its internal structures corresponding to
different decay channels before an actual decay has occurred.     \\

{\bf  7. Discussion and comments  }

    The representation of physical characteristics with the help
of Hilbert space spanned on Hamiltonian eigenfunctions caused that
sets of different physical states were represented by one
mathematical symbol, i.e., by one vector in the given Hilbert
space. All quantum-mechanical paradoxes have followed from such a
representation. The given mathematical framework has not been adequate to
describe the richness of real physical structures. To cope at
least partially with this problem two kinds of different physical
states (pure and mixed states) have been  introduced even if it was
not  practically possible to distinguish between them on
experimental basis.

   In the extended Hilbert space each actual physical state is represented by
a vector $\phi_{\tau,\kappa}(x_j)$ where $\kappa$ represents the
set of characteristics being conserved during the whole evolution.
The meaning of parameter $\tau$ may be illustrated, e.g., with the
help of a two-particle system; $\tau$ being the expectation value
of the operator $T$ defined by Eq. (\ref{tc}) or Eq. (\ref{th}) in
the two mentioned special cases. In the former case of conformal potential
(and similarly for all non-bound particle pairs)
$\tau = 0$ for the lowest possible mutual distance between both the
particles during the evolution of a given physical system.
Consequently, a numerical value of parameter $\tau$ characterizes the
instantaneous mutual distance of  given particles (expressed in time units).
During the evolution the change of  $\tau$ value is given by evolution
operator (\ref{ut}); $\tau$  increasing by  $t$.

  Hamiltonian eigenfunctions do not belong to the extended Hilbert
space and  do not represent any physical states in the case of the
extended model; the superposition
principle cannot be applied  to, either.   Any superposition of
vectors $\phi_{\tau,\kappa}(x_j)$ corresponds always to a statistical
mixture. And one should conclude that an actual shape of
$\psi(x_j)$ function has not
any direct physical meaning. The physical meaning may be attributed
to expectation values of corresponding operators, only. The same
evolution may be then described with the help of differently chosen
$\psi(x_j)$ functions (comp. Eq. (\ref{ini})), which may raise the question whether it is
necessary to use the representation Hilbert space defined over the
field of complex numbers in the extended model; even if the use of Hilbert
space may be still very helpful.

   As already stressed there is not any reason more to attribute the so much
discussed quantum paradoxes to the microscopic world. The extended
model seems to be based fully on particle picture of reality
(including photons representing the quanta of light). In
contradistinction to the standard quantum-mechanical model
the  extended model does not represent any closed physical theory;
any theory of everything has not  any place here. The model is
open for exploring still deeper characteristics of the microworld.
On the other side, it is not possible to state that a simple
particle picture represents fundamental and definite features of
matter reality; the question remaining open for further exploration.

  There are many other new
questions which have been opened now and which should be solved step by
step.  One of the central problems might concern the existence of time
operator and its definition;  time operator
being a function of $q_j$ and $p_j$. One should ask
whether the time is a basic quantity added to the space characteristics or a
quantity derived from the behavior of matter objects in the
usual three-dimensional space.

   And finally, it is necessary to stress that one should 
 respect again standard logical rules,
especially to take into account that any science or research cannot
bring  a decisive verification of our
ideas or hypotheses. The scientific methods are based on the falsification
approach, in principle. Since the only logical contradiction provides us with
a reliable response to our questions we may know only the untruth with
certainty, not the truth.  It is not, of course, possible to denote
as a truth what was already falsified in the past.

  On the other side, the falsification approach does not entitle us to deny
the existence of one truth about the world even if it must be perpetually
looked for in the region of possibilities left by the certainly known untruth.
  There is not more possible, either,  to argue on the basis of the
physical science that many-valued logic should be applied to natural
phenomena. There is not any reason to believe in the plurality of truth
concerning the world even if many modern philosophers try to convince us
about it, arguing often by the standard quantum-mechanical model and by
its paradoxes.    \\

 {\bf 8. Conclusion   }

    Concluding I should like to mention an actual position being claimed
for the extended model in the story of the physical science. It is possible
to say that in addition to Planck's discovery of the quantum structure of
energy
transfer and Rutherford's experiments they were Einstein and Bohr who
contributed mainly to the progress of the then physical research: Einstein by
the prediction of the photon \cite{ein4} and Bohr by formulating basic
postulates concerning the atom orbit structure \cite{bohr3}. And one can
also maintain that both these ideas belong still to fundamentals of the
contemporary physics, being based fully on the realistic view of matter
nature.

   However, it is necessary to admit that in other papers of theirs both these
physicists contributed fundamentally to that the realistic objective view
was changed to a formal mathematical description of physical phenomena:
Einstein by formulating the theory of special relativity \cite{ein5} and Bohr
by carrying out Copenhagen interpretation of the quantum-mechanical
mathematical model \cite{bohr7}. Even if it is possible to say that Einstein
personally returned to physical realism in papers concerning general
relativity it is evident that the common thinking in the whole century
has been fundamentally influenced by formal principles of special relativity
and mainly of quantum mechanics.

   As to the extended model it may be hardly regarded as
 a mere formal mathematical description of reality. It is
rather possible to say that through this model  the whole physical
story may return to the physical views hold in the beginning of this century
and characterized by the two formerly mentioned papers of Einstein
and Bohr.   \\

   I should like to appreciate highly numerous discussions about all related
 problems with my  colleagues Drs. J. Kr\'{a}sa and V. Kundr\'{a}t and to
 thank them also for many valuable comments to the  manuscript of this
paper.  \\   }



{\footnotesize
\begin{thebibliography}{99}
\bibitem{ein}
A. Einstein: Can quantum-mechanical description of physical reality be
considered complete?; Phys. Rev. 47 (1935), 777-80.
\bibitem{bohr}
N. Bohr: Can quantum-mechanical description of physical reality be
considered complete?;  \\  Phys. Rev. 48 (1935), 696-702.
\bibitem{phai}
 A.Mann, M.Revzen (eds): "The Dilemma of Einstein, Podolsky and
Rosen - 60 Years Later", Inst. of Phys., Publish. Techno House,
Bristol 1997.
\bibitem{gal}
A.Galindo, P.Pascual: Quantum mechanics I, Springer Verlag, 1990.
\bibitem{neu}
J. von Neumann:  Mathematische Grundlagen der Quantenmechanik; Springer
1932.
\bibitem{ros}
 D.M.Rosenbaum: Super Hilbert space and the quantum-mechanical time
operator;   \\  J. Math, Phys. 10 (1969), 1127-44.
\bibitem{lax}
P.D.Lax, R.S.Phillips: Scattering theory; Academic Press 1967.
\bibitem{alda}
V.Alda, V.Kundr\'{a}t, M.Lokaj\'{\i}\v{c}ek: Exponential decay and
irreversibility of decay and collision processes; Aplikace
matematiky 19 (1974), 307-15.
\bibitem{newt}
R.Newton: Quantum action-angle variables for harmonic oscillators;  \\
Ann. Phys. 124 (1980), 327-46.
\bibitem{lax2}
P.D.Lax, R.S.Phillips: Scattering theory for automorphic functions;
Princeton 1976.
\bibitem{jack}
R.Jackiw: Dynamical symmetry of the magnetic monopole; Ann. Phys.
129 (1980), 183-200.
\bibitem{dir}
P.A.M. Dirac: The quantum theory of the emission and absorption of
radiation;  \\  Proc. Roy. Soc. (London) A 114 (1927), 243-65.
\bibitem{lyn}
R. Lynch: The quantum phase problem, a critical review; Phys. Rep. 256
(1995), 367-437.
\bibitem{oza}
M. Ozawa: Phase operator problem and macroscopic extension of quantum
mechanics;  \\  Ann. Phys. 257 (1997), 65-83.
\bibitem{bell}
 J.S. Bell: On the Eimstein Podolsky Rosen paradox; Physics 1 (1964), 195-200.
\bibitem{asp}
 A. Aspect, P.Grangier, G.Roger: Experimental realization of
Einstein-Podolsky-Rosen-Bohm Gedankenexperiment: A new violation of Bell's
inequalities; Phys. Rev. Lett.  49 (1982), 91-4.
\bibitem{world}
A.Zeilinger et al.: Quantum information, teleportation, et al.;
 Physics World,  March 1998 (special issue), pp. 33-57.
\bibitem{lok3}
M.Lokaj\'{\i}\v{c}ek: Are teleportation and cryptography predicted by
quantum mechanics?; \\  http: //xxx.lanl.gov/quant-ph/9808019.
\bibitem{kly}
D.N.Klyshko: On the realization and interpretation of "quantum
teleportation"; \\  Phys. Lett. A 247 (1998), 261-6.
\bibitem{gold}
S.Goldstein: Quantum theory without observers; Physics Today,
March 1998, 42-6; April 1998,  38-42.
\bibitem{kazi}
M.Lokaj\'{\i}\v{c}ek: Limitations of standard quantum mechanics caused by
its current mathematical model and a way out; Proc. of XI Warsaw
Symposium on Elementary Particles "New Theories in Physics" (eds. Z.Ajduk,
S.Pokorski, A.Trautmann), Kazimierz, May 23-27, 1988, World Scientific,
Singapore, pp. 534-43.
\bibitem{lok1}
J.Kr\'{a}sa, V.Kundr\'{a}t, M.Lokaj\'{\i}\v{c}ek:   A new solution of
hidden-variable and measurement problems;  l.c. \cite{phai}, pp.87-90.
\bibitem{lok2}
M. Lokaj\'{\i}\v{c}ek: Locality problem, EPR experiments and Bell's
inequalities; \\  http: //xxx.lanl.gov/quant-ph/9808005.
\bibitem{belf}
F.J. Belifante: A survey of hidden-variable theories; Pergamon, Oxford 1973, 
p. 284.
\bibitem{kra1}
J.Kr\'{a}sa, J.Ji\v{r}i\v{c}ka, M.Lokaj\'{\i}\v{c}ek: Depolarization of light
by an imperfect polarizer;  \\  Phys. Rev. E 48 (1993), 3184-6.
\bibitem{kra2}
J.Kr\'{a}sa, M.Lokaj\'{\i}\v{c}ek, J.Ji\v{r}i\v{c}ka:
Transmittance of  laser beam through a pair of crossed polarizers;
Physics Letters A 186 (1994), 279-82.
\bibitem{pol}
  J.M.Movilla, G.Piquero, R.Martinez-Herrero, P.M.Mej\'{\i}as: Parametric
characterization of non-uniformly polarized beams; Optics Commun. 149
(1998), 230-4.
\bibitem{lok8}
M.V.Lokaj\'{\i}\v{c}ek: Resonances as unstable particles with
internal degrees of freedom; Czech. J. Phys. B 19 (1969), 1549-55.
\bibitem{lok9}
M.V.Lokaj\'{\i}\v{c}ek: Elastic partial-wave amplitude and poles of
higher order; Czech. J. Phys B 21 (1971), 823-7.
 \bibitem{ein4}
A.Einstein: \"{U}ber einen die Erzeugung und Verwandlung des Lichtes
betreffenden heuristischen Gesichtspunkt; Ann. Phys. 17 (1905), 132-48.
 \bibitem{bohr3}
N.Bohr: On the constitution of atom and molecules; Phil. Mag. 26 (1913),
1-25, 476, 857.
 \bibitem{ein5}
A.Einstein: Zur Elektrodynamik bewegter K\"{o}rper; Ann. Phys. 17
(1905), 891-921.
 \bibitem{bohr7}
N.Bohr: The quantum postulate and the development of atomic theory;
Nature 121 (1928), 580-90.
\end{thebibliography}  }
\end{document}